\documentclass[aps, prd, preprint, 11pt, superscriptaddress, nofootinbib]{revtex4-1}

\usepackage{fancyhdr}	
\usepackage{amsmath} \usepackage{amsfonts} \usepackage{amssymb}
\usepackage{wasysym}
\usepackage[dvipsnames]{xcolor}
\usepackage{graphicx}
\usepackage{multirow}
\usepackage{here}
\usepackage{epsfig}
\usepackage{soul}
\usepackage{hyperref}
\usepackage{float}

\renewcommand{\d}[1]{\mathinner{{\rm d}#1}}
\newcommand{\fn}[2]{\mathinner{#1\mathopen{\left(#2\right)}}}

\newcommand{\eq}[1]{Eq.~(\ref{#1})}
\newcommand{\eqs}[2]{Eqs.~(\ref{#1}) and (\ref{#2})}
\newcommand{\eqss}[3]{Eqs.~(\ref{#1}), (\ref{#2}) and (\ref{#3})}

\newcommand{\Mpc}{\mathinner{\mathrm{Mpc}}}
\newcommand{\MeV}{\mathinner{\mathrm{MeV}}}
\newcommand{\GeV}{\mathinner{\mathrm{GeV}}}
\newcommand{\TeV}{\mathinner{\mathrm{TeV}}}

\newcommand{\invMpc}{\mathinner{\mathrm{Mpc}^{-1}}}

\newcommand{\fref}[1]{Figure~\ref{#1}}
\newcommand{\sref}[1]{Section~\ref{#1}}

\begin{document}

\title{CMB Spectral Distortion Constraints on Thermal Inflation}

\author{Kihyun Cho}
\email{cho$_$physics@kaist.ac.kr}
\affiliation{Department of Physics, KAIST,
Daejeon 34141, Republic of Korea}

\author{Sungwook E. Hong}
\affiliation{Korea Astronomy and Space Science Institute,
Daejeon 34055, Republic of Korea}

\author{Ewan D. Stewart}
\affiliation{Department of Physics, KAIST,
Daejeon 34141, Republic of Korea}

\author{Heeseung Zoe}
\email{heezoe@dgist.ac.kr}
\affiliation{School of Undergraduate Studies,
College of Transdisciplinary Studies,
Daegu Gyeongbuk Institute of Science and Technology (DGIST),
Daegu 42988, Republic of Korea}
\date{\today}

\begin{abstract}
Thermal inflation is a second epoch of exponential expansion at typical energy scales $V^{1/4} \sim 10^{6 \sim 8}\GeV$. 
If the usual primordial inflation is followed by thermal inflation, the primordial power spectrum is only modestly redshifted on large scales, but strongly suppressed on scales smaller than the horizon size at the beginning of thermal inflation, $k > k_{\rm b} = a_{\rm b} H_{\rm b}$.
We calculate the spectral distortion of the cosmic microwave background generated by the dissipation of acoustic waves in this context.
For $k_{\rm b} \ll 10^3 \invMpc$, thermal inflation results in a large suppression of the $\mu$-distortion amplitude, predicting that it falls well below the standard value of $\mu \simeq 2\times 10^{-8}$.
Thus, future spectral distortion experiments, similar to PIXIE, can place new limits on the thermal inflation scenario, constraining $k_{\rm b} \gtrsim 10^3 \invMpc$ if $\mu \simeq 2\times 10^{-8}$ were found. 
\end{abstract}


\maketitle

\section{Introduction}

Thermal inflation \cite{Lyth:1995hj,Lyth:1995ka,Yamamoto:1985mb,Yamamoto:1985rd,Enqvist:1985kz,Bertolami:1987xb,Ellis:1986nn,Ellis:1989ii,Randall:1994fr} is a brief low-energy inflation phase motivated to resolve the moduli problem in supersymmetric cosmology \cite{Coughlan:1983ci,Banks:1993en,deCarlos:1993jw}. 
Moduli, scalar fields with Planckian vacuum expectation values, are dangerous because they decay late disturbing Big Bang nucleosynthesis (BBN).
This moduli problem can be resolved when a flaton, an unstable flat direction which is generic in the supersymmetric theories, is held at the origin by thermal effects.
Its vacuum energy drives an inflationary phase diluting the moduli to safe abundance. 
Thermal inflation also resolves the gravitino problem \cite{Coughlan:1983ci,Banks:1993en,deCarlos:1993jw}, provides a mechanism for baryogenesis \cite{Stewart:1996ai,Jeong:2004hy,Kawasaki:2006py,Felder:2007iz,Kim:2008yu,Lazarides:1985ja,Yamamoto:1986jw,Mohapatra:1986dg} and implements a viable axion/axino dark matter cosmology \cite{Kim:2008yu,Moxhay:1984am}. 

\begin{figure}[hbt]
\centering
\includegraphics[width=1\textwidth]{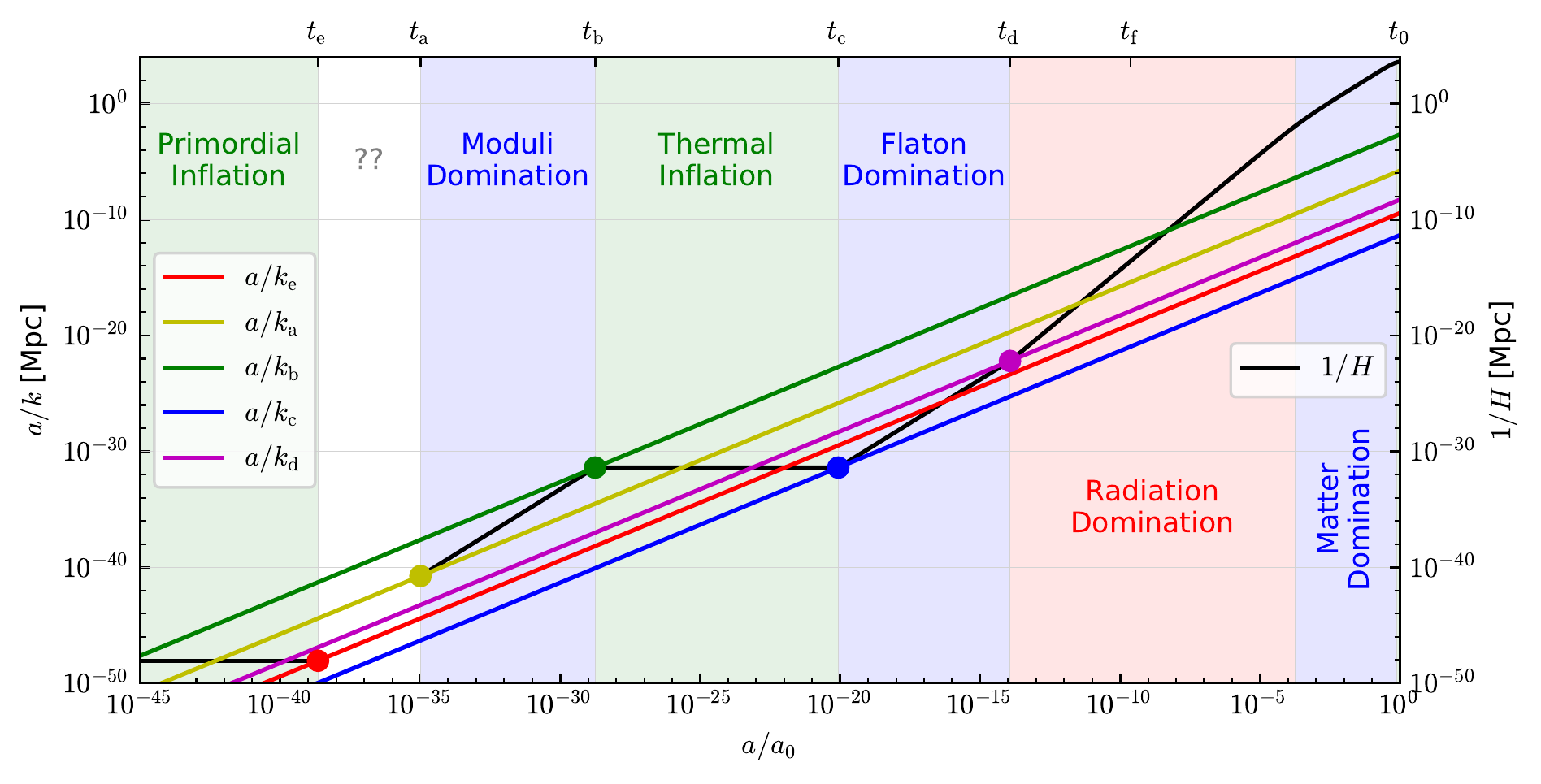}
\caption{
History of the observable universe with thermal inflation.
Shaded regions: epochs dominated by {\color{red}radiation-like}, {\color{blue}matter-like}, or {\color{OliveGreen}vacuum-like} energy components.
Solid lines: the evolution of $k$-modes (colors) and the Hubble radius (black).
Subscripts denote the time when each $k$-mode is at the Hubble radius.
}
\label{fig:kb}
\end{figure}

\fref{fig:kb} summarizes thermal inflation cosmology. 
Primordial inflation \cite{gliner1,gliner2,Guth:1980zm,Linde:1981mu,Albrecht:1982wi}, as the first observable inflation epoch, generates scale-invariant perturbations which are observed in the cosmic microwave background (CMB) and grow to form today's large-scale structure (LSS). 
After an unknown post-inflation period, moduli matter starts dominating when the Hubble parameter of the universe drops below the moduli mass scale.
Thermal inflation occurs {\it after} moduli generation but before BBN, solving the moduli problem, and when the temperature drops below the flaton's mass scale, thermal inflation ends giving rise to a period of flaton matter domination. 
The flaton decay results in the usual radiation domination of BBN.

During these post-primordial-inflation epochs, cosmological perturbations on large scales, seen in CMB and LSS observations, remain outside the horizon but those on small scales enter and re-exit the horizon, affecting the shape and amplitude of the small-scale power spectrum \cite{Hong:2015oqa}. 
Thus the primordial power spectrum is expected to be preserved and only modestly redshifted on large scales, but strongly suppressed on scales smaller than the horizon size at the beginning of thermal inflation. 
Hence, thermal inflation scenarios may be tested by probes of small-scale power such as observations of ultracompact minihalos or primordial black holes \cite{Carr:1975qj, Josan:2009qn, Bringmann:2011ut}, the lensing dispersion of SNIa \cite{Ben-Dayan:2013eza, Ben-Dayan:2014iya, Ben-Dayan:2015zha}, the 21cm hydrogen line at or prior to the epoch of reionization \cite{Cooray:2006km, Mao:2008ug} or, as we focus on in this paper, CMB distortions \cite{Chluba:2015bqa, Chluba:2011hw, Chluba:2012gq, Chluba:2012we}.

Measurements with COBE/FIRAS proved that the CMB spectrum is extremely close to that of a blackbody with temperature $T_\gamma = 2.726 \pm 0.001 {\rm K}$ and spectral distortions limited to $\Delta I / I \lesssim 5\times 10^{-5}$ \cite{Fixsen:1996nj,Mather:1993ij}. 
However, there are many possible sources of energy release that could affect the CMB energy spectrum and lead to spectral distortions, including the dissipation of primordial density perturbations \cite{barrow,Chluba:2012we,Chluba:2013dna,Chluba:2012gq,Daly,Ganc:2012ae,Hu:1994bz,Pajer:2013oca,SZ1970,Clesse:2014pna,Emami:2015xqa,Dimastrogiovanni:2016aul,Chluba:2016aln}, reionization and structure formation \cite{Cen:1998hc,Hu:1993tc,Miniati:2000iu,Oh:2003sa,Refregier:2000xz,SZ1972,Zhang:2003nr,Hill:2015tqa}, decaying or annihilating particles \cite{Chluba:2013wsa,Chluba:2013pya,Hu:1993gc,McDonald:2000bk,Sarkar:1984tt}, cosmic strings \cite{Ostriker:1986xc,Tashiro:2012nb,Tashiro:2012nv,Tashiro:2012pp}, primordial black holes \cite{Carr:2009jm,Pani:2013hpa,Tashiro:2008sf}, small-scale magnetic fields \cite{Jedamzik:1999bm,Kunze:2013uja,Sethi:2004pe,Chluba:2015lpa}, the adiabatic cooling of matter \cite{Chluba:2011hw, Ali-Haimoud:2015pwa}, cosmological recombination \cite{Dubrovich1975, Dubrovich1997, RubinoMartin:2006ug,Chluba:2006xa,SC2009,Chluba:2015gta} and gravitino decay \cite{Dimastrogiovanni:2015wvk}. 
At $z \gg 2 \times 10^6$, any energy release can be efficiently thermalized by Compton or double Compton scattering and Bremsstrahlung, restoring the energy spectrum to a blackbody spectrum \cite{SZmu, Hu:1993tc, Burigana}. 
The contribution of these interactions to thermalization changes as the universe expands and cools, determining the type of CMB spectral distortion. 
At $ 3 \times 10^5 \lesssim z \lesssim 2 \times 10^6 $, double Compton and Bremsstrahlung processes are gradually reduced and energy or photons injected to the CMB would only be redistributed over frequency, predominantly by Compton scattering. 
In this epoch, electrons and photons are kept in kinetic equilibrium by Compton scattering and a nonzero chemical potential $\mu$ is formed, with the associated distortion being called $\mu$-distortion \cite{SZmu}. 
At $z \lesssim 10^4$, Compton scattering events between photons and electrons become very inefficient and photons diffuse only a little in energy. 
In this regime, energy release causes a distortion that is called (Compton) $y$-distortion, a signature that is also known in connection with the Sunyaev-Zeldovich effect of galaxy clusters \cite{SZ1969}.
Between these redshifts, at $10^4\lesssim z  \lesssim 3 \times 10^5$, Compton scatterings become inefficient in redistributing photons over frequency and the distortion can be described as a sum of $\mu$ and $y$-distortions with a smaller residual (non-$\mu$/non-$y$) distortion, called $r$-distortion \cite{Chluba:2013pya, Chluba:2015bqa}.

The primary concern of this paper is the dissipation of primordial density fluctuations by Silk damping of acoustic waves, due to the shear viscosity in the baryon-photon fluid \cite{SZ1970, Daly, barrow, Chluba:2012we, Hu:1994bz, Chluba:2012gq}. 
In thermal inflation scenarios, the power spectrum is suppressed on small scales \cite{Hong:2015oqa}, so the energy release due to the dissipation process should also be reduced.
This suggests that with future CMB spectral distortion measurements, the parameter space of thermal inflation could be constrained. 
While COBE/FIRAS has already imposed tight upper bounds on the main distortion parameters, $|\mu| \lesssim 9 \times 10^{-5}$ and $|y| \lesssim 1.5 \times 10^{-5}$ (95\% c.l.), future concepts like the Primordial Inflation Explorer (PIXIE) \cite{Kogut:2011xw} could have a sensitivity of $|\mu| \sim 10^{-8}$ and $ |y| \sim 10^{-9}$ (68\% c.l.) \cite{Chluba:2013pya}. 
Hence, CMB spectral distortion seems a promising avenue towards constraining the thermal inflation scenario.

The paper is structured as follows: we discuss the curvature perturbations generated by primordial inflation followed by thermal inflation in \sref{PS}. 
We specify the primordial power spectra for broad class of simple inflation models and combine it with the thermal inflation transfer function. 
In \sref{cmb_distortion}, we estimate the CMB distortions due to the dissipation of acoustic waves for a scenario of primordial inflation followed by thermal inflation (thermal inflation scenario) and compare it with a standard scenario of primordial inflation followed by radiation domination (standard scenario).
We find that thermal inflation could be constrained with PIXIE-type CMB distortion observations. 
In \sref{dis}, we review the main results and discuss future work.

\section{Power spectrum of the thermal inflation scenario}
\label{PS}

\subsection{General formalism for a primordial power spectrum modified by thermal inflation}
\label{ps_multiple}
In this section, we describe our general framework for a curvature power spectrum generated by primordial inflation and modified by thermal inflation.
The power spectrum of single-field slow-roll primordial inflation is given by
\begin{equation}\label{srps}
\fn{\mathcal{P}_{\rm pri}}{\frac{k}{k_{\rm e}}} = \left. \left( \frac{H}{2 \pi} \frac{H}{\dot{\phi}}\right)^2 \right|_{\frac{aH}{a_{\rm e} H_{\rm e}}=\frac{k}{k_{\rm e}}},
\end{equation}
where $k_{\rm e} = a_{\rm e} H_{\rm e}$ and the subscript `e' denotes the end of primordial inflation. For thermal inflation beginning at $t = t_{\rm b}$ after primordial inflation, the amplitude of modes with wavelength $k < k_{\rm b} = a_{\rm b} H_{\rm b}$ is not altered, as they remain outside the horizon.
However, modes with $k > k_{\rm b}$ enter the horizon before, and may re-exit during, thermal inflation so that their amplitudes are significantly affected. 
These effects can be expressed in terms of a transfer function $\fn{\mathcal{T}}{k/k_{\rm b}}$ \cite{Hong:2015oqa}. 
The observable power spectrum is then
\begin{equation}\label{tips}
\fn{\mathcal{P}_{\rm obs}}{\frac{k}{k_*},\frac{k_{\rm e}}{k_*},\frac{k_{\rm b}}{k_*}} 
= \fn{\mathcal{P}_{\rm pri}}{\frac{k}{k_{\rm e}}}
\fn{\mathcal{T}^2}{\frac{k}{k_{\rm b}}},
\end{equation}
where $k_*$ is a convenient observational reference scale,  
\begin{equation}
k_* = 0.05 \invMpc 
\end{equation}
which we take to be the Planck pivot scale \cite{Ade:2015lrj}.

The parameters $k_{\rm e}/k_*$ and $k_{\rm b}/k_*$ are determined by the deflation histories after primordial inflation, and since the beginning of thermal inflation, respectively.
For arbitrary times $t_x$ and $t_y$,
\begin{equation}\label{Nxy}
\ln \frac{k_y}{k_x} = \mathcal{N}_{xy} 
\end{equation}
where 
\begin{equation}
\mathcal{N}_{xy} \equiv \ln \frac{a_yH_y}{a_xH_x}
\end{equation}
is the net inflation from $t_x$ to $t_y$ and $k_x = a_x H_x$ is the comoving scale that crosses the horizon at $t = t_x$.
When the epoch between $t_x$ and $t_y$ is described by a single energy component, with energy density $\rho \propto a^{-3(1+\omega)}$, 
\begin{align}
\mathcal{N}_{xy} 
& = \ln \frac{a_y H_y}{a_x H_x} \\
& = - \frac{1}{2} \left( 1 + 3 \omega \right) N_{xy} \\
& = - \left( \frac{1+3\omega}{3+3\omega} \right) \ln\frac{H_x}{H_y} \label{Nxyomega} \, ,
\end{align}
where $N_{xy} \equiv \ln(a_y/a_x)$ is the number of $e$-folds of expansion from $t_x$ to $t_y$.

Defining $t_{\rm f}$ by the condition 
\begin{equation} 
T_{\rm f} \equiv 1 \MeV
\end{equation}
we separate the early universe from the observationally tested late universe, such that our parameters 
$k_{\rm e}/k_*$ and $k_{\rm b}/k_*$ can be expressed as 
\begin{align}\label{Ne*def}
\ln \frac{k_{\rm e}}{k_*} & =  \mathcal{N}_{\rm *e} = -\mathcal{N}_{\rm ef} + \mathcal{N}_{\rm *f} \\
\ln \frac{k_{\rm b}}{k_*} & = \mathcal{N}_{\rm *b} = -\mathcal{N}_{\rm bf} + \mathcal{N}_{\rm *f} \,,
\end{align}
where
\begin{eqnarray}
\mathcal{N}_{\rm *f} && =  \ln \frac{a_{\rm f}}{a_0} + \ln H_{\rm f} + \ln \frac{a_0}{k_*} \\
&& =  \ln \frac{ \fn{g_{\rm *s}^{1/3}}{T_0} T_0 }{ \fn{g_{\rm *s}^{1/3}}{T_{\rm f}} T_{\rm f} } + \ln \frac{ \pi \fn{g^{1/2}_*}{T_{\rm f}} T_{\rm f}^2 }{ 3 \sqrt{10}\, m_{\rm Pl} } + \ln \frac{a_0}{k_*} \\
&& \simeq  12.4 \, ,\label{est_kfk*}
\end{eqnarray}
where $g_*(T)$ and $g_{\rm *s}(T)$ are the effective number of degrees of freedom for calculating energy density and entropy, respectively \cite{Kolb:1990vq}. 
Thus, to specify the observable power spectrum, \eq{tips}, we only need the primordial power spectrum, $\fn{\mathcal{P}_{\rm pri}}{k/k_{\rm e}}$, the thermal inflation transfer function, $\fn{\mathcal{T}^2}{k/k_{\rm b}}$, and the early universe deflation history parameters, $\mathcal{N}_{\rm ef}$ and $\mathcal{N}_{\rm bf}$.

\subsection{Primordial power spectrum}
\label{simple_assume}

In many simple inflation models, the spectral index of primordial inflation takes the form  
\begin{equation} \label{n_1overN}
\frac{d \ln \mathcal{P}_{\rm pri} }{\d \ln k} \left( \frac{k}{k_{\rm e}} \right)
= - \frac{c}{\mathcal{N}_{\rm e}} 
+ \fn{\mathcal{O}}{ \frac{1}{\mathcal{N}^2_{\rm e}} } \, ,
\end{equation}
and hence
\begin{equation}
\fn{\mathcal{P}_{\rm pri}}{\frac{k}{k_{\rm e}}} 
\simeq A \,\mathcal{N}_{\rm e}^c
\end{equation}
where
\begin{equation}
\mathcal{N}_{\rm e} = 
\left. \ln \frac{a_{\rm e} H_{\rm e}}{aH} \right|_{aH=k}
= \ln \frac{k_{\rm e}}{k}
\end{equation}
is the amount of inflation from horizon exit ($aH = k$) to the end of inflation, and $A$ and $c$ are constants that depend on the model of inflation.

Assuming $k_* \ll k_{\rm b}$, so that the effect of the transfer function in \eq{tips} is negligible (see \sref{sec:transfer}), we can use observations to fix
\begin{align}
\left. \mathcal{P}_{\rm obs} \right|_{k=k_*} & = A_* \\
\left. \frac{\d \ln \mathcal{P}_{\rm obs} }{\d \ln k} \right|_{k=k_*} & =  n_* - 1 \, .
\end{align}
From now on,
we use $A_* = 2.21 \times 10^{-9}$ and $n_* = 0.96$ from Planck 2015 \cite{Ade:2015lrj}.
Expressing $\mathcal{N}_{\rm e}$ in terms of the observational reference scale $k_*$
\begin{equation} 
\mathcal{N}_{\rm e} = \mathcal{N}_{\rm *e} - \ln \frac{k}{k_*}~,
\end{equation}
we get\footnote{Equivalently, $\mathcal{P}_{\rm pri} \simeq A_* \left( \frac{k}{k_*} \right)^{(n_*-1)\sum_{j=0}^{\infty} \frac{1}{j+1} \left(  \frac{\ln \frac{k}{k_*}}{\mathcal{N}_{\rm *e}}  \right)^j}$ 
with $\mathcal{N}_{\rm *e} = \frac{n_*-1}{n'_*}$ where $n'_* \equiv \left. \frac{dn}{d\ln k} \right|_{k=k_*}$, but note that truncating this series would not be correct as the logs are large for the scales we are considering. }
\begin{equation} \label{ps_est}
\fn{\mathcal{P}_{\rm pri}}{\frac{k}{k_*},\frac{k_{\rm e}}{k_*}} 
\simeq A_* \left( 1 - \frac{1}{\mathcal{N}_{\rm *e}} \ln \frac{k}{k_*} \right)^{ (1-n_*) \mathcal{N}_{\rm *e} } \, ,
\end{equation}
with the post-inflation history encoded in 
\begin{equation} 
\ln \frac{k_{\rm e}}{k_*} = \mathcal{N}_{\rm *e} = -\mathcal{N}_{\rm ef} + \mathcal{N}_{\rm *f}~.
\end{equation}

\subsection{Deflation histories}

We define $t_{\rm a}$ by the condition 
\begin{equation} 
H_{\rm a} = 1 \TeV
\end{equation}
to separate the early universe history into two eras, one associated with primordial inflation and its aftermath, and the other with thermal inflation. 
The nature of the post-primordial-inflation epoch, $t_{\rm e} \leq t \leq t_{\rm a}$, is unknown, but it begins at the end of primordial inflation when $H = H_{\rm e}$ with $H_{\rm a} \leq H_{\rm e} \lesssim 2 \times 10^{14} \GeV$ \cite{Ade:2015tva, Ade:2015lrj} and ends when $H = H_{\rm a}$, and we assume an equation of state with $0 \leq p \leq \rho/3$, giving
\begin{equation} \label{Nea}
- 13.0 \lesssim
- \frac{1}{2} \ln \frac{H_{\rm e}}{H_{\rm a}}
\leq \mathcal{N}_{\rm ea} \leq
-\frac{1}{3} \ln \frac{H_{\rm e}}{H_{\rm a}} \leq 0.
\end{equation} 

We consider two scenarios for the history $t_{\rm a} \leq t \leq t_{\rm f}$: a thermal inflation scenario (superscript TI) and a standard scenario (superscript S).

\subsubsection{Thermal inflation scenario}
The thermal inflation-related era begins with moduli matter domination ($t_{\rm a} \leq t \leq t_{\rm b}$), which provided the original motivation for thermal inflation. 
This is followed by thermal inflation ($t_{\rm b} \leq t \leq t_{\rm c}$), with $H_{\rm c} = H_{\rm b}$, which dilutes the moduli to safe abundances. 
When the escape of the flaton field from its finite temperature potential ends the thermal inflation at $t = t_{\rm c}$, the inflationary potential energy is converted to flaton oscillations, 
giving rise to an epoch of flaton-matter domination ($t_{\rm c} \leq t \leq t_{\rm d}$). 
Finally, the flaton field decays at $t = t_{\rm d} < t_{\rm f}$ giving rise to the radiation domination epoch of the late universe, see \fref{fig:kb}. 

Thus, in the thermal inflation scenario, 
we have
\begin{equation} \label{HaHf}
\ln \frac{H_{\rm a}}{H_{\rm f}} 
= \ln \frac{ 3 \sqrt{10}\, m_{\rm Pl} H_{\rm a} }{ \pi \fn{g^{1/2}_*}{T_{\rm f}} T_{\rm f}^2 }
\simeq 63.0
\end{equation}
\begin{equation} \label{HaHb}
\ln \frac{H_{\rm a}}{H_{\rm b}} 
= \ln \frac{\sqrt{3} m_{\rm Pl} H_{\rm a} }{ V^{1/2}_0}
\simeq 17.6 - 2 \ln \frac{V_0^{1/4}}{10^7 \GeV}
\end{equation}
and
\begin{equation} \label{HdHf}
\ln \frac{H_{\rm d}}{H_{\rm f}} = 2 \ln \frac{ \fn{g^{1/4}_*}{T_{\rm d}} T_{\rm d} }{ \fn{g^{1/4}_*}{T_{\rm f}} T_{\rm f} }
\simeq 15.2 + 2 \ln \left[ \left( \frac{ \fn{g_*}{T_{\rm d}} }{ 10^2 } \right)^{1/4} \left( \frac{ T_{\rm d} }{ \GeV } \right) \right] \, ,
\end{equation}
which, using \eq{Nxyomega}, give
\begin{equation} \label{Naf_thermal}
\begin{split}
\mathcal{N}^{\rm TI}_{\rm af} &  = N_{\rm bc} 
- \frac{1}{3} \ln \frac{H_{\rm a}}{H_{\rm f}} 
- \frac{1}{6} \ln \frac{H_{\rm d}}{H_{\rm f}}  \\
& \simeq N_{\rm bc} - 23.5 - \frac{1}{3} \ln \left[ \left( \frac{ \fn{g_*}{T_{\rm d}} }{ 10^2 } \right)^{1/4} \left( \frac{ T_{\rm d} }{ \GeV } \right) \right] \, , 
\end{split}
\end{equation}

\begin{equation} \label{Nab}
\mathcal{N}^{\rm TI}_{\rm ab} 
= - \frac{1}{3} \ln \frac{H_{\rm a}}{H_{\rm b}}  \simeq -5.9 + \frac{2}{3} \ln \left( \frac{V_0^{1/4}}{10^7\GeV} \right) \, ,
\end{equation}
and
\begin{equation} \label{NTI_bf}
\begin{split}
\mathcal{N}_{\rm bf}^{\rm TI} 
& = N_{\rm bc} 
- \frac{1}{3} \ln \frac{H_{\rm a}}{H_{\rm f}} 
+ \frac{1}{3} \ln \frac{H_{\rm a}}{H_{\rm b}} 
- \frac{1}{6} \ln \frac{H_{\rm d}}{H_{\rm f}} \\
& \simeq  N_{\rm bc} -18.6 
- \frac{1}{3} \ln \left[
\left( \frac{ \fn{g_*}{T_{\rm d}} }{ 10^2 } \right)^{1/4} 
\left( \frac{ T_{\rm d} }{ \GeV } \right) 
\left( \frac{ V_0^{1/4} }{ 10^7 \GeV } \right)^2
\right] \, ,
\end{split}
\end{equation}
where $V_0$ is the thermal inflationary potential energy density,
$N_{\rm bc}$ is the number of $e$-folds of thermal inflation, 
and  $T_{\rm d}$ is the temperature after flaton decay.

\subsubsection{Standard scenario}
In this scenario, we assume radiation domination from $t_{\rm a}$ to $t_{\rm f}$, and so, using  \eq{HaHf}, 
\begin{equation} \label{NRD_af}
\mathcal{N}^{\rm S}_{\rm af} =  - \frac{1}{2} \ln \frac{H_{\rm a}}{H_{\rm f}} \simeq - 31.5 \, .
\end{equation}

\subsection{Thermal inflation transfer function $\fn{\mathcal{T}}{k/k_{\rm b}}$}
\label{sec:transfer}

\begin{figure}
\centering
\includegraphics[width=0.75\textwidth]{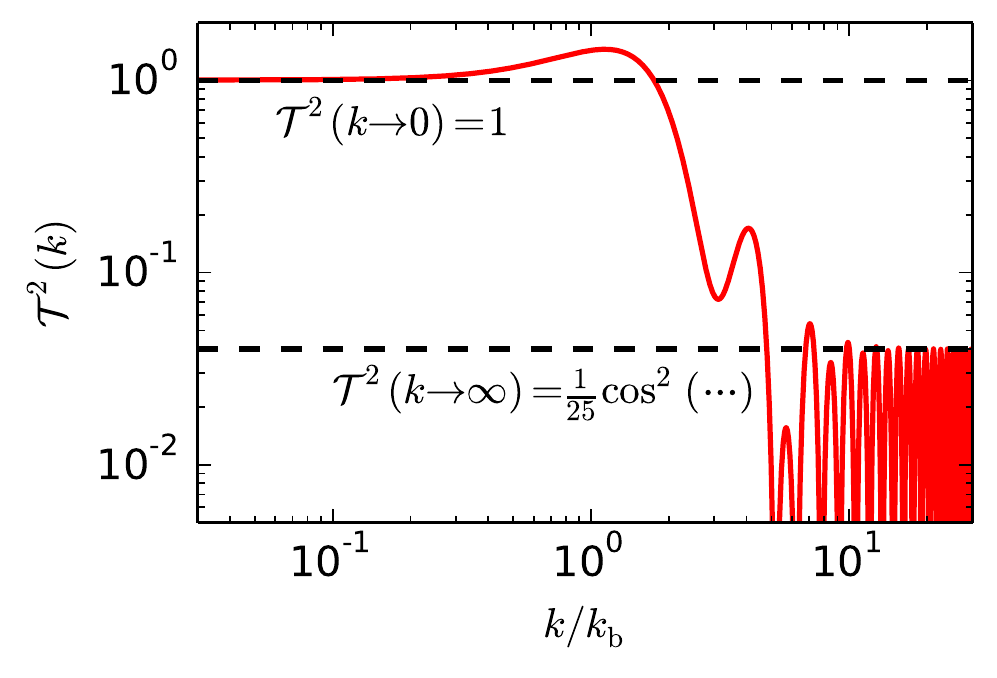}
\caption{ \label{fig:transfer}
Transfer function of \eq{transfer} induced by thermal inflation  \cite{Hong:2015oqa}. 
The first peak at $k \simeq 1.13 k_{\rm b}$ has amplitude $\mathcal{T}^2 \simeq 1.45$, the first dip at $k \simeq 3.11 k_{\rm b}$ has amplitude $\mathcal{T}^2 \simeq 0.07$ and the power spectrum is suppressed by a factor $50$ on small scales.
}
\end{figure}

Primordial curvature perturbations with $k > k_\mathrm{b}$ enter the horizon during moduli matter domination to become
perturbations in the dominant moduli and subdominant radiation components, which evolve until they rexit the horizon during
thermal inflation. Thermal inflation ends when the drop in temperature triggers a phase transition, converting the radiation perturbations into curvature perturbations. The effect of this process on the primordial spectrum is described by the transfer function derived in \cite{Hong:2015oqa},
\begin{multline}\label{transfer}
\fn{\mathcal{T}}{\frac{k}{k_{\rm b}}} = \cos \left[ \left( \frac{k}{k_{\rm b}} \right) \int_0^\infty \frac{\d\alpha}{\sqrt{\alpha(2+\alpha^3)}} \right] \\
+ 6 \left( \frac{k}{k_{\rm b}} \right) \int_0^\infty \frac{\d\gamma}{\gamma^3} \int_0^\gamma \d\beta \left( \frac{\beta}{2+\beta^3}\right)^{3/2}
\sin \left[ \left( \frac{k}{k_{\rm b}} \right) \int_\gamma^\infty \frac{\d\alpha}{\sqrt{\alpha(2+\alpha^3)}} \right]~.
\end{multline}
See \fref{fig:transfer}.

The asymptotic behavior of the transfer function is
\begin{equation}\label{kappazero}
\fn{\mathcal{T}}{\frac{k}{k_{\rm b}}} \to
\begin{cases}
& \begin{displaystyle} 1 + \nu_0 \left( \frac{k}{k_{\rm b}} \right)^2 + \mathcal{O}\left[\left( \frac{k}{k_{\rm b}} \right)^4\right] \textrm{ as } \frac{k}{k_{\rm b}}  \to 0 \end{displaystyle} \\
& \begin{displaystyle} - \frac{1}{5} \cos \left[ \nu_1 \left( \frac{k}{k_{\rm b}} \right) \right] + \mathcal{O}\left[\left( \frac{k}{k_{\rm b}} \right)^{-n} \right] \textrm{ as } \frac{k}{k_{\rm b}} \to \infty  \end{displaystyle}.
\end{cases}
\end{equation}
where 
\begin{align}
\nu_0 &\equiv \int_0^\infty \d\alpha \left( \frac{\alpha}{2+\alpha^3} \right)^{3/2}
\simeq 0.3622 \\
\nu_1 &\equiv \int_0^\infty \frac{\d\alpha}{\sqrt{\alpha(2+\alpha^3)}}
\simeq 2.2258 ~.
\end{align}

\eq{tips} gives
\begin{equation}\label{n*}
\frac{\d{\ln \mathcal{P}_{\rm obs}}}{\d{\ln k}} = \frac{\d{\ln \mathcal{P}_{\rm pri}}}{\d{\ln k}}
+ 2 \frac{\d{\ln \mathcal{T}}}{\d{\ln k}}
\end{equation}
and, on large scales, \eq{kappazero} yields 
\begin{equation}
2 \frac{\d{\ln \mathcal{T}}}{\d{\ln k}} \sim 4 \nu_0 \left( \frac{k}{k_{\rm b}} \right)^2~.
\end{equation}
Thus, if 
\begin{equation}
\left(\frac{k}{k_{\rm b}} \right)^2 \ll \frac{1-n_*}{4 \nu_0} \sim 3 \times 10^{-2}
\end{equation}
i.e. 
$k_{\rm b} \gtrsim 1 \Mpc$, then, 
the transfer function does not contribute to the spectral index at $k\simeq k_*$.

\subsection{Summary}
\label{ps_form}

In the standard scenario, primordial inflation with a spectral index of the form of \eq{n_1overN} is followed by radiation domination between $H_{\rm a} \equiv 1 \TeV$ and $T_{\rm f} \equiv 1\MeV$. The power spectrum is
\begin{equation}\label{ps_typical_est}
\fn{\mathcal{P}^{\rm S}_{\rm obs}}{\frac{k}{k_*},\frac{k_{\rm e}^{\rm S}}{k_*}} 
\simeq A_* \left[ 1 - \frac{1}{\mathcal{N}^{\rm S}_{\rm *e}} \ln \left( \frac{k}{k_*} \right)  \right]^{(1-n_*) \mathcal{N}^{\rm S}_{\rm *e}},
\end{equation}
with 
\begin{equation}
\begin{split}
\ln \frac{k_{\rm e}^{\rm S}}{k_*} & = \mathcal{N}_{\rm *e}^{\rm S}  = -\mathcal{N}_{\rm ea} - \mathcal{N}_{\rm af}^{\rm S} + \mathcal{N}_{\rm *f} \\
& \simeq 44 -\mathcal{N}_{\rm ea} 
\end{split}
\end{equation}
estimated to lie in the range
\begin{equation}\label{NSrange}
44\lesssim \mathcal{N}_{\rm *e}^{\rm S} \lesssim 57\, 
\end{equation}
where we have used \eqss{est_kfk*}{Nea}{NRD_af}.
The case of a pure power law primordial spectrum corresponds to taking $\mathcal{N}_{\rm ea} \to - \infty$, in which case $\mathcal{N}_{\rm *e} \to \infty$ and \eq{ps_typical_est} reduces to
\begin{equation}
\fn{\mathcal{P}^{\rm S}_{\rm obs}}{\frac{k}{k_*}}  = A_* \left( \frac{k}{k_*} \right)^{n_*-1}~.
\end{equation}

In the thermal inflation scenario, primordial inflation with a spectral index of the form of \eq{n_1overN} is followed by thermal inflation. The power spectrum is
\begin{equation}\label{ps_thermal_est}
\fn{\mathcal{P}^{\rm TI}_{\rm obs}}{\frac{k}{k_*},\frac{k_{\rm e}^{\rm TI}}{k_*},\frac{k_{\rm b}}{k_*}} 
\simeq A_* \left[ 1 - \frac{1}{\mathcal{N}^{\rm TI}_{\rm *e}} \ln \left( \frac{k}{k_*} \right)  \right]^{(1-n_*) \mathcal{N}^{\rm TI}_{\rm *e}}
\fn{\mathcal{T}^2}{\frac{k}{k_{\rm b}}},
\end{equation}
with $\mathcal{T}(k/k_{\rm b})$ given in \eq{transfer} and
\begin{equation}
\begin{split}
\ln \frac{k_{\rm e}^{\rm TI}}{k_*} = \mathcal{N}_{\rm *e}^{\rm TI} 
& = \mathcal{N}_{\rm *b}^{\rm TI} - \mathcal{N}_{\rm ea} - \mathcal{N}_{\rm ab}^{\rm TI} \\
& \simeq 6 + \ln \frac{k_{\rm b}}{k_*} - \mathcal{N}_{\rm ea} 
 - \frac{2}{3} \ln \left( \frac{V_0^{1/4}}{10^7\GeV} \right)
\end{split}\label{ke_TI}
\end{equation}
estimated to lie in the range
\begin{equation}\label{NTIrange}
13 + \ln \left( \frac{k_{\rm b}}{10^3 \invMpc} \right)
\lesssim \mathcal{N}_{\rm *e}^{\rm TI}  \lesssim 
32 + \ln \left( \frac{k_{\rm b}}{10^3 \invMpc} \right) \, 
\end{equation}
where we have used \eqss{Nea}{HaHb}{Nab}, and assumed $10^5\GeV \leq V_0^{1/4} \leq 10^9\GeV$. 

The characteristic scale $k_{\rm b}$ depends on the amount of inflation during the thermal inflation epoch
\begin{equation}
\ln \frac{k_{\rm b}}{k_*} = \mathcal{N}_{\rm *f} - \mathcal{N}^{\rm TI}_{\rm bf} \simeq 31 - N_{\rm bc} 
+ \frac{1}{3} \ln 
\left[ 
\left( \frac{\fn{g_*}{T_{\rm d}}}{10^2}  \right)^{1/4}
\left( \frac{T_{\rm d}}{\GeV}  \right)
\left( \frac{V_0^{1/4}}{10^7 \GeV}  \right)^2
\right]\, 
\end{equation}
or, equivalently, 
\begin{equation} \label{kb}
k_{\rm b} \simeq 3 \times 10^3\, \invMpc \left( \frac{e^{20}}{e^{N_{\rm bc}}} \right)
\left( \frac{\fn{g_*}{T_{\rm d}}}{10^2}  \right)^{1/{12}}
\left( \frac{T_{\rm d}}{\GeV}  \right)^{1/3}
\left( \frac{V_0^{1/4}}{10^7 \GeV}  \right)^{2/3}
  \, . 
\end{equation}
While $N_{\rm bc} \sim 10$ is typical in thermal inflation scenarios and single thermal inflation has $N_{\rm bc} \lesssim 15$, multiple thermal inflation in quite natural \cite{Lyth:1995ka, Felder:2007iz, Kim:2008yu, Choi:2012ye} so that there is no theoretical upper bound on $N_{\rm bc}$, and $k_{\rm b}$ can be small enough to leave observable signatures in CMB spectral distortions.

In the standard scenario, the primordial inflationary parameter
\begin{equation}
c = \mathcal{N}_{\rm *e} \left( 1 - n_* \right)
\end{equation}
of Eq.~(15) can be reasonably well determined by measuring $n_*$ due to the relatively narrow range of $\mathcal{N}_{\rm *e}^{\rm S}$ in Eq.~(41). However, in the thermal inflation scenario, $\mathcal{N}_{\rm *e}^{\rm TI}$ can take a wide range of values leaving $c$ undetermined by $n_*$ alone, requiring either $n'_*$ to be measured to determine $c$ via
\begin{equation}
c = - \frac{ \left( n_* - 1 \right)^2 }{ n'_* }
\end{equation}
or $k_{\rm b}$, and hence $\mathcal{N}_{\rm *e}^{\rm TI}$, to be sufficiently constrained.

\begin{figure}
\centering
\includegraphics[width=0.75\textwidth]{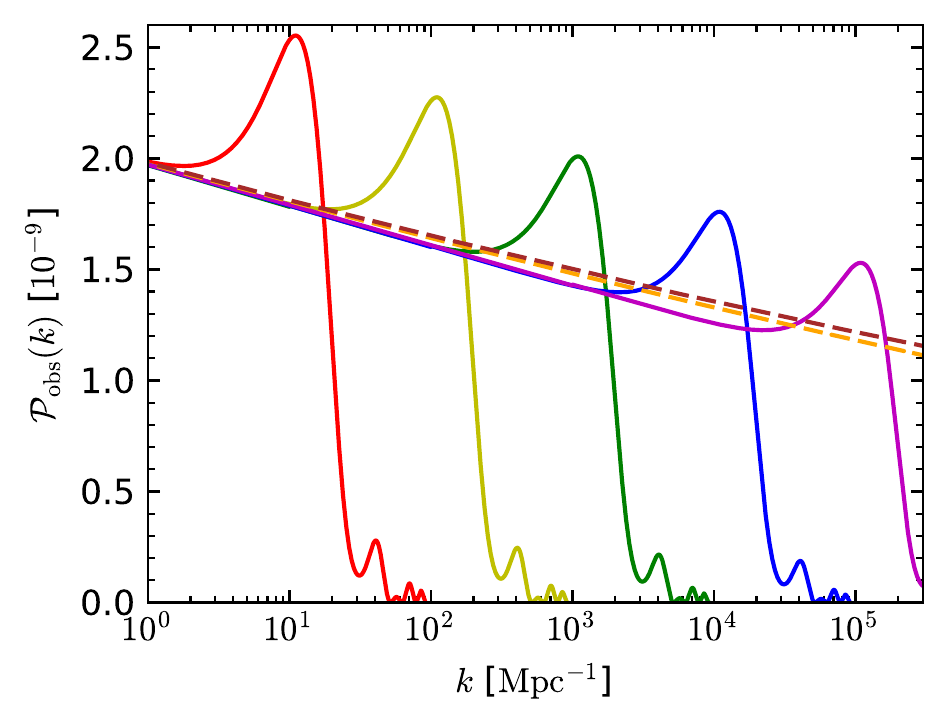}
\caption{Power spectra of thermal inflation and standard scenarios.
Solid lines: thermal inflation scenarios with $\mathcal{N}_{\rm ea} + \frac{2}{3} \ln \left( \frac{V_0^{1/4}}{10^7\GeV} \right)  = -7$ and $k_{\rm b}/\invMpc = {\color{red}10}$, $\color{Goldenrod}10^2$, $\color{OliveGreen}10^3$, $\color{blue}10^4$, $\color{violet}10^5$.
Dashed lines: standard scenarios with $\mathcal{N}_{\rm ea} = {\color{orange}0}$, $\color{brown}-13$, $- \infty$.
}\label{fig:Pk}
\end{figure}

The power spectra of the thermal inflation and standard scenarios are shown in \fref{fig:Pk}.
The slope of power spectrum in the thermal inflation scenario at $k \ll k_{\rm b}$ is steeper than in the standard scenario with the same $\mathcal{N}_{\rm ea}$ because of the redshifting effect of thermal inflation. See \eqs{NSrange}{NTIrange}. 
The power spectrum then starts to turn up at $k \simeq 0.27 k_{\rm b}$, reaching a maximum at $k \simeq 1.06 k_{\rm b}$, after which it drops steeply to  a local minimum at $k \simeq 3.12 k_{\rm b}$, and then starts oscillating.

\section{CMB distortions in the thermal inflation scenario}
\label{cmb_distortion}
Large-scale observations of the CMB and LSS only give the mild constraint on thermal inflation of $k_{\rm b} \gtrsim 1 \invMpc$ \cite{Hong:2015oqa}.
However, CMB spectral distortions can probe smaller scales, and hence give a stronger constraint on thermal inflation.

CMB spectral distortions can be calculated using a Green-function method \cite{Chluba:2013vsa, Chluba:2015hma}. 
The spectral distortion at a given frequency $\Delta I_\nu$ is estimated from the heating rate $\d (Q/\rho_\gamma)/ \d z$ by
\begin{equation}
\Delta I_\nu \simeq \int \fn{G_{\rm th}}{\nu, z'}  \frac{\d(Q/\rho_\gamma)}{\d z'} \d z'
\end{equation}
where $ \fn{G_{\rm th}}{\nu, z'} $ includes the relevant thermalisation physics, which is independent
of the energy release scenario.  

\subsection{Heating rate}
In the tight coupling approximation, which is relevant for the period long before recombination \cite{Hu:1995en}, 
the heating rate due to the dissipation of acoustic waves with adiabatic initial conditions is 
\begin{equation}\label{heatingrate}
\frac{\d (Q/\rho_\gamma)}{\d z} 
\simeq 4A_{\rm h}^2 \int \d \ln k \fn{\mathcal{P}_{\rm obs}}{k} W_{\rm s}(k) W_{\rm D}(k) \, ,
\end{equation}
where $A_{\rm h} \simeq 0.9$ and
\begin{align}
W_{\rm s}(k) & = \sin^2 \left (k r_{\rm s} \right) \\
W_{\rm D}(k) & = \frac{6}{1+z} \frac{k^2}{k_{\rm D}^2} \exp \left( -\frac{2 k^2}{k_{\rm D}^2} \right) \, ,
\end{align}
are the window functions related to the sound horizon \cite{Chluba:2012gq, Chluba:2012we, Chluba:2015bqa} 
\begin{equation}
r_{\rm s} \simeq 2.7 \times 10^5 \Mpc (1+z)^{-1}
\end{equation}
and the damping scale \cite{Chluba:2012gq, Chluba:2012we, Chluba:2015bqa}
\begin{equation}\label{damping_scale}
k_{\rm D} \simeq 4.0\times 10^{-6} \invMpc (1+z)^{3/2}~.
\end{equation}

For $k_{\rm D} r_{\rm s} \gg 2\pi$, i.e. $z \gg 30$, $W_{\rm s}(k)$ oscillates rapidly and can be approximated as $1/2$. 
Also, $W_{\rm D}(k)$ is sharply peaked at $k \sim k_{\rm D}$. 
Therefore, \eq{heatingrate} becomes
\begin{equation}\label{heatingrate_approx2}
(1+z)\frac{\d (Q/\rho_\gamma)}{\d z} \sim \fn{\mathcal{P}_{\rm obs}}{k_{\rm D}(z)} \, .
\end{equation}
Thus, if a certain feature (e.g. maximum or minimum) occurs in $\mathcal{P}_{\rm obs}(k)$ at $k$, one can estimate the redshift $z$ at which a similar feature will occur in $(1+z) \d (Q/\rho_\gamma) / \d z$ by inverting \eq{damping_scale} to give 
\begin{equation}
z  \sim (4.0 \times 10^3) \left( \frac{k}{\invMpc} \right)^{2/3}
\end{equation}

\begin{figure}[tpb]
\centering
\includegraphics[width=0.85\textwidth]{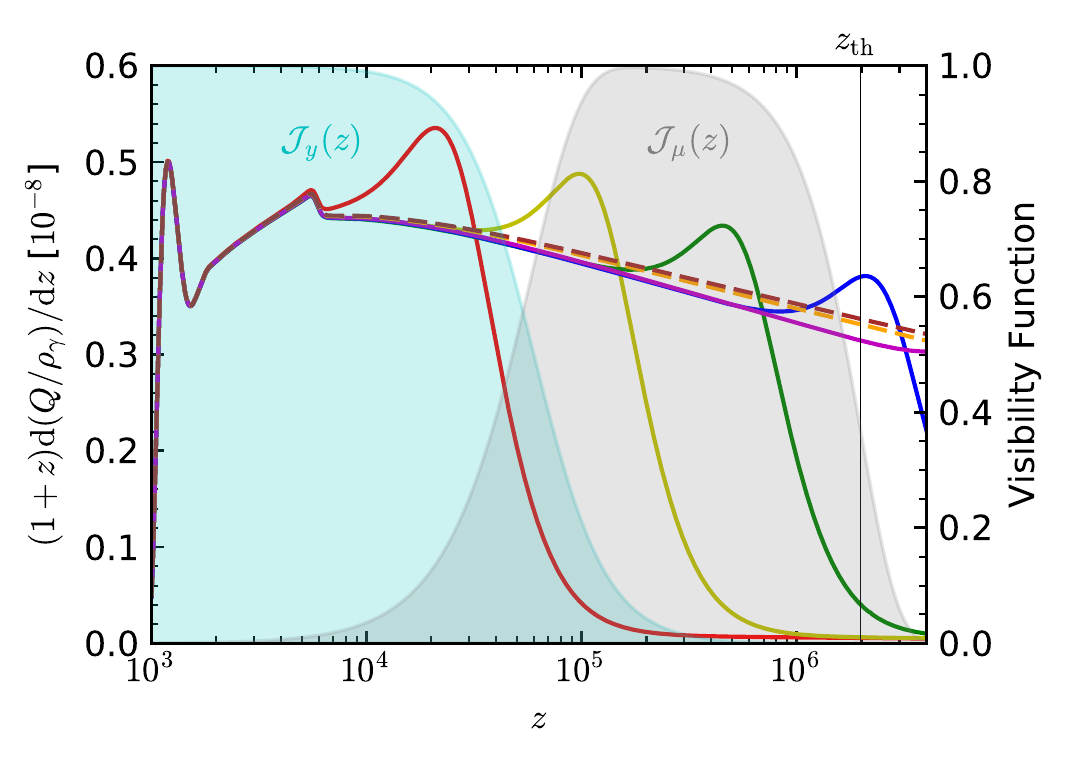}
\caption{ \label{fig:heating}
Heating rates of thermal inflation and standard scenarios.
Solid lines: thermal inflation scenario with $\mathcal{N}_{\rm ea} + \frac{2}{3} \ln \left( \frac{V_0^{1/4}}{10^7\GeV} \right)  = -7$ and $k_{\rm b} = {\color{red}10}$, $\color{Goldenrod}10^2$, $\color{OliveGreen}10^3$, $\color{blue}10^4$, ${\color{violet}10^5}~\invMpc$.
Dashed lines: standard scenario with $\mathcal{N}_{\rm ea} = {\color{orange}0}$, $\color{brown}-13$, $- \infty$.
Shaded regions: visibility functions for $y$-distortion ($\mathcal{J}_y$) and $\mu$-distortion ($\mathcal{J}_\mu$).
}
\end{figure}

\fref{fig:heating} shows the heating rate of thermal inflation scenarios as a function of redshift, which  have a similar form to their corresponding power spectra, dipping at
\begin{equation}
z_\frown \simeq (4.5 \times 10^3) \left( \frac{k_{\rm b}}{\invMpc} \right)^{2/3} 
\end{equation}
and rising to a maximum at 
\begin{equation}
z_\smile \simeq (1.3 \times 10^3) \left( \frac{k_{\rm b}}{\invMpc} \right)^{2/3} 
\end{equation}
before dropping off sharply at large $z$.

\subsection{$\mu$- and $y$-distortions}

Once the heating rate is found, the CMB spectral distortion $\Delta I_\nu$, the change in the original blackbody spectrum after the energy release over the frequency domain, can be expressed in terms of the temperature shift, $y$ and $\mu$ contributions
\begin{equation}\label{distortions}
\Delta I_\nu \approx \frac{\Delta T}{T} \fn{G}{\nu} + y \fn{Y_\mathrm{SZ}}{\nu} + \mu \fn{M_\mathrm{SZ}}{\nu}.
\end{equation}

The first term in \eq{distortions} describes a temperature shift that changes the blackbody spectrum by an amount proportional to
\begin{equation}
\fn{G}{\nu} = T \frac{\partial \fn{B}{\nu}}{\partial T}
\end{equation}
where
\begin{equation}
\fn{B}{\nu} = \frac{2h \nu^3}{c^2 (e^x - 1)}
\end{equation}
and 
\begin{equation}
x = \frac{h \nu}{k_B T} \, .
\end{equation}

The second term in \eq{distortions} is the $y$-distortion \cite{SZ1969}, with spectral shape 
\begin{equation}
\fn{Y_{\rm SZ}}{\nu} \simeq T \frac{\partial \fn{B}{\nu}}{\partial T} \left( x \fn{\coth}{x}-4  \right)
\end{equation}
and $y$-parameter 
\begin{equation}\label{y_visibility_function}
y \simeq \frac{1}{4} \int_{z_{\rm rec}}^\infty \fn{\mathcal{J}_y}{z'} 
\frac{\d{\left( Q/\rho_\gamma \right)}}{\d z'} \d z' ,
\end{equation}
where $z_{\rm rec}\approx 10^3$. 
The $y$-visibility function can be estimated as \cite{Chluba:2013vsa, Chluba:2016bvg}
\begin{equation}
\fn{\mathcal{J}_y}{z} \simeq \left[ 1 + \left( \frac{1 + z}{1 + z_y}   \right)^{2.58}   \right]^{-1} \, ,
\end{equation}
where $z_y \approx 6 \times 10^4$ (see the cyan shaded region in \fref{fig:heating}).

The third term in \eq{distortions} is the $\mu$-distortion \cite{SZmu}, with
\begin{equation}
\fn{M_\mathrm{SZ}}{\nu} \simeq T \frac{\partial \fn{B}{\nu}}{\partial T} \left( 0.4561 - \frac{1}{x}  \right)
\end{equation}
and the $\mu$-parameter is
\begin{equation}\label{J_mu}
\mu \simeq 1.401 \int_0^\infty \fn{\mathcal{J}_\mu}{z'} \frac{\d{\left( Q/\rho_\gamma\right)}}{\d z'} \d z' \, .
\end{equation}
A simple approximation of the $\mu$-visibility function is given in \cite{Chluba:2016bvg}:
\begin{equation}
\fn{\mathcal{J}_\mu}{z} \simeq \left[1 - \exp \left( - \left[ \frac{1 + z}{1 +z_\mu}   \right]^{1.88}  \right)\right]
\,\exp\left( -\left[\frac{1 + z}{1 + z_{\rm th}}\right]^{2.5}\right) \, ,
\end{equation}
where $z_\mu \approx 5.8 \times 10^4$ and $z_{\rm th} \approx 1.98 \times 10^6$ (see the gray shaded region in \fref{fig:heating}).
The last factor accounts for the efficiency of thermalization, which becomes very high above the thermalization redshift ($z_{\rm th}$) \cite{Burigana, Hu:1992dc}.
However, in this paper, we calculate the $\mu$-value with the full thermalization Green's function \cite{Chluba:2013vsa, Chluba:2015hma}, whose numerical code has been developed by Jens Chluba, with applying the distortion eigenmode method described in \cite{Chluba:2013pya}.

\begin{figure}[H]
\centering
\includegraphics[width=0.85\textwidth]{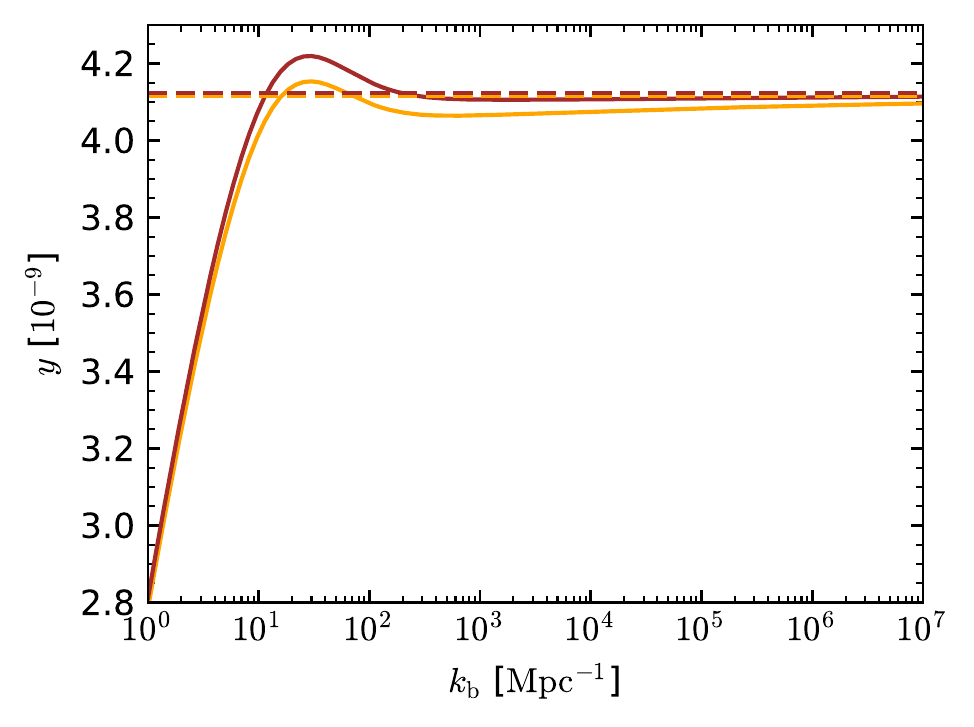}
\caption{$y$-distortions in thermal inflation scenarios with $\mathcal{N}_{\rm ea} + \frac{2}{3} \ln \left( \frac{V_0^{1/4}}{10^7\GeV} \right)  = {\color{orange}0}$ and $\color{brown}-13$ as a function of $k_{\rm b}$.
The dashed lines show the corresponding standard scenarios.}\label{fig:kby}
\end{figure}

\begin{figure}[h]
\centering
\includegraphics[width=0.85\textwidth]{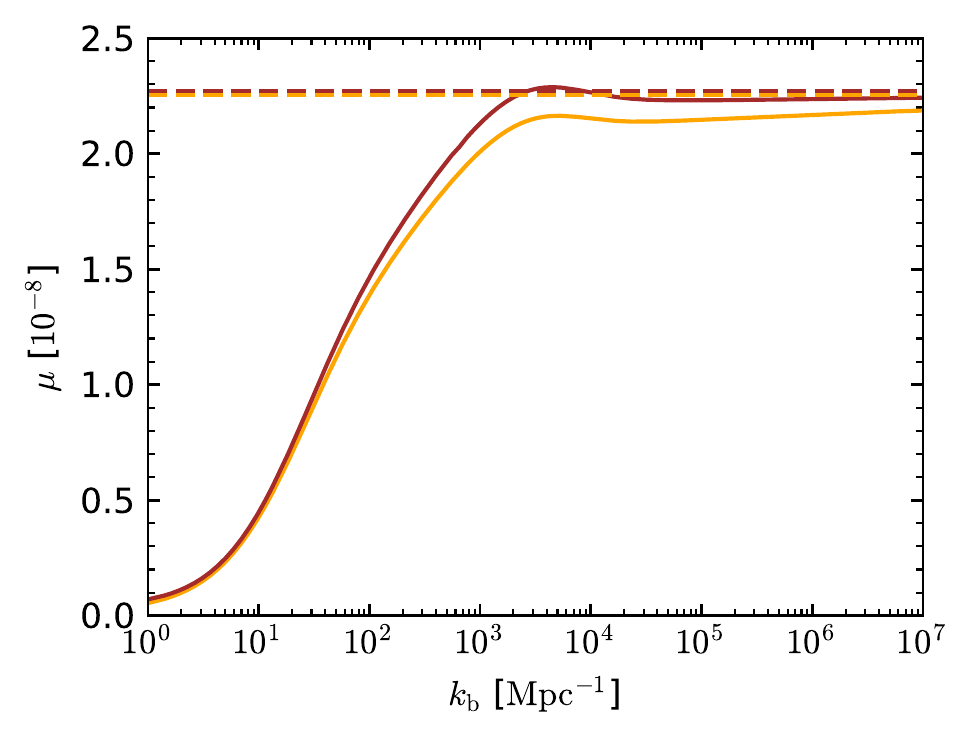}
\caption{$\mu$-distortions in thermal inflation scenarios with $\mathcal{N}_{\rm ea} + \frac{2}{3} \ln \left( \frac{V_0^{1/4}}{10^7\GeV} \right)  = {\color{orange}0}$ and $\color{brown}-13$ as a function of $k_{\rm b}$.
The dashed lines show the corresponding standard scenarios.}\label{fig:kbmu}
\end{figure}

\begin{figure}[h]
\centering
\includegraphics[width=0.85\textwidth]{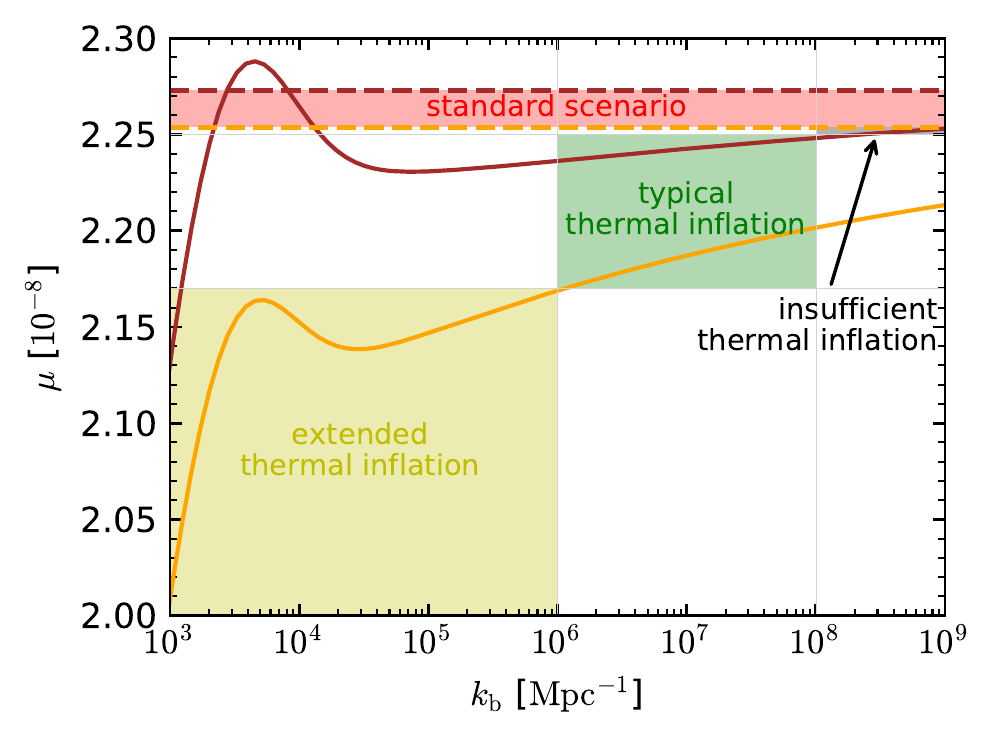}
\caption{Zoom-in plot of \fref{fig:kbmu} showing how to interpret precision measurements of the $\mu$-distortions.}
\label{fig:kbmufuture}
\end{figure}

Figures~\ref{fig:kby}, \ref{fig:kbmu} and \ref{fig:kbmufuture} show the $y$- and $\mu$-values for the thermal inflation and standard scenarios as a function of $k_{\rm b}$. 
Their form can be understood by considering the overlap of the heating rate and visibility functions in \fref{fig:heating}.
The graphs of $y$ and $\mu$ have peaks at $k_{\rm b} \simeq 30 \invMpc$ and $k_{\rm b} \simeq 3\times 10^3 \invMpc$  respectively, drop off sharply at smaller $k_{\rm b}$, and dip to minima at $k_{\rm b} \simeq 3 \times 10^2 \invMpc$ and $k_{\rm b} \simeq 3\times 10^4 \invMpc$, respectively, before gradually asymptoting to the standard scenario values as the redshifting effect diminishes at large $k_{\rm b}$ \cite{Chluba:2012gq, Chluba:2013pya, Cabass:2016giw, Chluba:2016bvg}.

In calculating the $\mu$-distortion, the small correction due to the cooling of baryons relative to photons $\Delta \mu = -0.334 \times 10^{-8}$ \cite{Chluba:2011hw, Chluba:2016bvg} was neglected. 
As CMB photons heat up the non-relativistic plasma of baryons by Compton scattering, we need to consider such energy extraction from the photons to the baryons \cite{Chluba:2011hw}. 
The same process applies in the thermal inflation scenario and the corrections need to be considered. 
By including $\Delta \mu = -0.334 \times 10^{-8}$, the value of $\mu$ in the thermal inflation scenario can become negative for $k_{\rm b} \lesssim 7 \invMpc$.

By considering the proposed $y$-sensitivity of PIXIE ($ |y| \sim 10^{-9}$; 68\% c.l.), the $y$-parameter would be expected to be detected by PIXIE for thermal inflation scenarios with $k_{\rm b} \gtrsim 1 \invMpc$.
However, unless $k_{\rm b} \lesssim 10 \invMpc$, it is unlikely that the thermal inflation scenario can be constrained using the $y$-distortion in the near future.
On the other hand, the thermal inflation scenario can be constrained by the proposed $\mu$-sensitivity of PIXIE ($|\mu| \sim 10^{-8}$; 68\% c.l.) in the following ways.
If the $\mu$-distortion is detected at the level of the standard scenario ($\mu \simeq 2 \times 10^{-8}$), then the parameter space of thermal inflation will be constrained to $k_{\rm b} \gtrsim 10^3 \invMpc$.
However, if the $\mu$-distortion is observed to be less than $2 \times 10^{-8}$, thermal inflation with $k_{\rm b} < 10^3 \invMpc$ can be an attractive way of explaining such a small value of the $\mu$-distortion while remaining fully consistent with existing constraints at large scales. 

When CMB distortion observations with precision $ |\mu| \sim 10^{-10}$ become available, it would be possible to test the standard scenario and typical thermal inflation scenarios, see \fref{fig:kbmufuture}.
\begin{itemize}
\item Values of $\mu$ in the red band would strongly favor the standard scenario and rule out all but fine tuned thermal inflation scenarios with $k_{\rm b} \sim 5 \times 10^3 \invMpc$.
\item Values of $\mu$ in the green region would confirm the prediction of a typical thermal inflation scenario with $k_{\rm b} \sim 10^6$ to $10^8 \invMpc$ but could also be explained by fine tuned standard scenarios with altered primordial spectrum.
\item Values of $\mu$ in the yellow region could be explained by a multiple thermal inflation scenario with more than typical number of $e$-folds but also by modified standard scenarios with a dipping primordial spectrum \cite{Nakama:2017ohe}.
\end{itemize}

\section{Discussion} 
\label{dis}
In this paper, we calculated the power spectrum of primordial inflation followed by thermal inflation, see \eq{ps_thermal_est} and \fref{fig:Pk}.
It differs from the power spectrum of primordial inflation followed by radiation domination in the following two aspects.
First, the power spectrum in the thermal inflation scenario is slightly enhanced and then suppressed by a factor of 50 on scales smaller than $k_{\rm b}$, the horizon scale at the beginning of thermal inflation, see \eq{transfer} and \fref{fig:transfer}.
Second, it is redshifted by an amount that can be parameterized in terms of $k_{\rm b}$, see \eq{ke_TI}.   
Hence, the net effects of thermal inflation on small-scale power spectrum can be effectively parameterized by a single parameter $k_{\rm b}$.     

We showed how future observations of CMB spectral distortions such as PIXIE can constrain thermal inflation.
We focussed on the difference between thermal inflation and standard scenarios in the prediction of $\mu$-distortions generated by dissipation of acoustic waves, which can be detected by PIXIE-like observations with $|\mu| \sim 10^{-8}$ (68\% c.l.).
For $k_{\rm b} \lesssim 10^3 \invMpc$, there is a large suppression of the $\mu$-distortion (see \fref{fig:kbmu}).  
If future observations do not detect a $\mu$-distortion at the level of $2 \times 10^{-8}$, multiple thermal inflation can be an attractive explanation while remaining fully consistent with existing constraints at large scales.
In contrast, if $\mu\simeq 2 \times 10^{-8}$ is found, thermal inflation will be constrained to $k_{\rm b} \gtrsim 10^3 \invMpc$.

We leave further comparisons between thermal inflation and other scenarios having strong suppression of power spectrum at small scales, including warm dark matter scenarios, as future works. 
It may be possible to distinguish them with small scale observations such as the residual distortions \cite{Chluba:2013pya, Chluba:2015bqa} and the substructure of galaxies \cite{Moore:1999nt}.

\begin{acknowledgements} 
The authors thank Jens Chluba for discussions and suggestions at various stages of this work, and Kyungjin Ahn, Donghui Jeong and Subodh Patil for useful discussions. 
\end{acknowledgements}

\bibliographystyle{apsrev4-1}
\bibliography{CMBdistortions_TIP}

\end{document}